# A Two-State Picture of Water and the Funnel of Life

Lars G.M. Pettersson[*]

*Department of Physics, AlbaNova University Center, Stockholm University, S-106 91 Stockholm, Sweden*

[*]Corresponding author: Lars G.M. Pettersson, lgm@fysik.su.se

**Abstract**

Here I show that experimental and simulation data on liquid water using vibrational (infrared and Raman) and X-ray (absorption and emission) spectroscopies, as well as recent data from X-ray scattering, are fully consistent with a two-state picture of water. At ambient conditions there are fluctuations between a dominating high-density liquid (HDL) and a low-density form (LDL). These are related to the two forms of amorphous ice at very low temperature, high-density amorphous (HDA) and low-density amorphous (LDA), which interconvert in a first-order-like transition. This transition line is assumed to continue into the so-called "No-man's land" as a liquid-liquid transition and terminate in a critical point with very large fluctuations between the two liquid forms. These fluctuations extend in a funnel-like region up to ambient temperatures and pressures and give water its unusual properties which are fundamental to life. With this picture we find simple, intuitive explanations of the anomalous properties of water, such as the density maximum at 4 °C, why ice floats, and why the compressibility and heat capacity grow as the liquid is cooled. We summarize by noting that in this picture, water is not a complicated liquid, but two normal liquids with a complicated relationship.

I. **INTRODUCTION**

Our planet Earth is often called "The Blue Planet" due to the large part of its surface covered by oceans, which appear blue from space. Water is considered a prerequisite for life and it thus becomes important not only to know both the extent and quality of the water resources that are available, but also to understand the properties of water in order to efficiently produce the fresh water that is necessary for an increasing population of humans on Earth. A recent illustration from the US Geological Survey[1], based on the estimates by Shiklomanov[2], shows the total amount of available water in, on and above the Earth as a sphere of about 1,385 kilometers in diameter, which is quite small in comparison with that (12,742 kilometers) of the Earth. However, the dominating fraction is salty ocean water and, when this is excluded, the sphere representing the remainder (trapped in glaciers and ice caps, groundwater, swamp water, rivers, and lakes) has shrunk to a sphere only 272.8 kilometers in diameter. Most of this fresh water is inaccessible. When only easily accessible surface water in lakes and rivers is included, the diameter of the corresponding sphere is a mere 56.2 kilometers. Clearly, fresh water is a limited resource and we will need to develop efficient approaches to large-scale desalination to supply the needs for fresh water for the growing human population, for agriculture, for industry and to abate desertification in view of global climate change. To achieve this, a profound understanding of water's properties is a prerequisite. The overarching question then is: How well do we understand water?

Water is often referred to as "the most anomalous liquid"[3] due to the number of physical and chemical properties in which it deviates from the behavior of normal simple liquids[4]. Examples of these anomalies include that the solid (ice) is less dense than the liquid and that the density of the liquid is maximum at 4 °C. These properties of water make aquatic life in temperate zones possible, since lakes and oceans freeze over from the top, providing an insulating layer, while the temperature at the bottom remains at 4 °C. The specific heat of the liquid (4.18 J/g·K), *i.e.* how much energy is needed to change the temperature by one degree, is about twice that of the solid and significantly higher than for the simple liquid ethanol (2.44 J/g·K). This provides the basis for water as an efficient medium to redistribute energy over the Earth, as exemplified by the Gulf Stream. The specific heat ($C_P$) depends on fluctuations in entropy, while the compressibility at constant temperature ($\kappa_T$) of the liquid depends on fluctuations in the density. At higher temperatures both decrease in magnitude with decreasing temperature, as expected for any normal liquid as thermal motion slows down. However, water exhibits a minimum at 35 °C ($C_P$) and 46 °C ($\kappa_T$) after which both $C_P$ and $\kappa_T$ seem to go towards infinity near a temperature of -45 °C upon further cooling. This behavior is clearly anomalous in that fluctuations thus dramatically *increase* as the liquid is cooled below 0 °C[5-9].

Another example of anomalous behavior is given by the boiling point of water at +100 °C which, if extrapolated based on decreasing molecular mass along the group 6A dihydrides ($H_2Te$, $H_2Se$, $H_2S$ and $H_2O$), should rather be closer to -66 °C. The boiling point of water is thus ~160 °C higher than expected from comparable liquids (Fig. 1). Also the melting point, when extrapolated in the same manner, is around 90 °C higher than expected[4]. As final examples we note the unusually high surface tension and self-diffusivity, *i.e.* molecular mobility, where the latter is observed to *increase* with applied pressure when the liquid is cooled below the minimum in compressibility at 46 °C. Clearly, the origin of the anomalous properties must be sought in how the water molecules interact with each other, and what types of local arrangements they can form under different conditions.

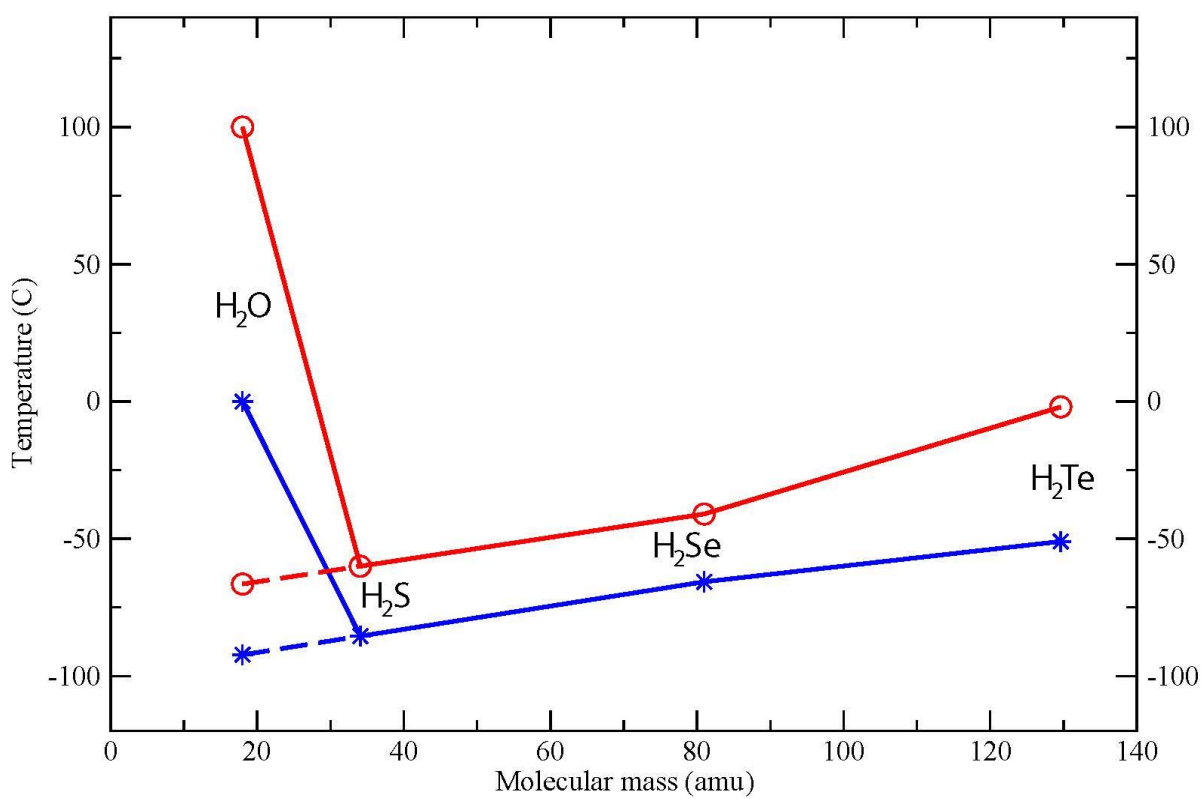

**Figure 1.** Melting point (blue line, stars) and boiling point (red line, circles) of the group 6A dihydrides as function of molecular mass. Dashed curves show straight line extrapolations to $H_2O$ using the trend from $H_2Se$ to $H_2S$. $H_2Po$ (mass 211, melting point -36 °C, boiling point 37 °C) is not shown in order to better focus on the region around water. Temperatures in °C.

A two-state picture of water provides a simple and intuitive understanding of the origin of water's many anomalous properties by considering the liquid under ambient temperatures and pressures as balancing between two local forms. One form favors order, through maximizing hydrogen-bond (H-bond) formation, which limits the number of neighbors and leads to a local low-density liquid (LDL) environment. The other form favors disorder, squeezing the molecules tighter (close-packing) by breaking or distorting H-bonds, thus leading to a local high-density liquid (HDL)[10-25]. The two forms are clearly incompatible and at ambient conditions the HDL form dominates, but with local fluctuations into the LDL form as it is lower in energy.

The two forms of the liquid (LDL, HDL) have been connected to the two forms of amorphous solid water, *i.e.* low-density amorphous (LDA) and high-density amorphous (HDA) ice, which differ in density by ~20% and exhibit a first-order-like transition between them[26-27]. By heating HDA to above -163 °C (110 K), *i.e.* into the ultra-viscous regime, it has recently been possible to follow the transition between the two forms (HDL to LDL) using wide-angle X-ray scattering (WAXS) and simultaneously measuring diffusion through X-ray photon correlation spectroscopy (XPCS) [28]. Through this combination of techniques the transition was shown to proceed from HDA into HDL, that subsequently converted into LDL, *i.e.* a liquid-liquid transition[28]. Thus, the existence of, and transformation between, the hypothesized two forms of liquid water seems clear, albeit only in deeply supercooled and pressurized form.

A coexistence line between two phases, *e.g.*, gas and liquid, may terminate in a critical point where fluctuations grow beyond bounds. Beyond the coexistence line and critical point extends a funnel-like region of fluctuations between the two forms. The fluctuations become less intense the farther away one moves into the one-phase region beyond the critical point where one can no longer distinguish the two phases. A remaining question thus concerns whether the coexistence line between the two liquids, HDL and LDL, also is terminated by a critical point, which could explain the seeming divergence of compressibility and heat capacity as water is deeply supercooled[16,23,29]. Experimentally, the probable location of such a critical point around 800 bar[8] and significantly below the temperature of homogeneous ice nucleation, makes it very challenging to unambiguously determine its existence. However, also far away from such a critical point, one would expect local fluctuations between the two forms. This would lead to two-state behavior with fluctuations on some characteristic length- and time-scale. The magnitude of such fluctuations would depend both on temperature and pressure, as these two thermodynamic quantities determine the distance from the hypothesized liquid-liquid critical point (LLCP). If such structural fluctuations can be shown to exist also for ambient water it would provide very simple and intuitive explanations of all of water's anomalous properties.

Molecular dynamics (MD) computer simulations have contributed significantly to the present understanding of the liquid, based on both classical force-field interactions[30] and quantum mechanical techniques, *e.g.*, references 31-34. Impressive agreement with experiment has been obtained for many properties, such as the radial distribution functions, molecular self-diffusion, proton transport, and many more. Even the challenging problem of the difference in density between simulated liquid and ice has been resolved using advanced simulation techniques[31-32]. These are properties that depend on averaging, while for properties that depend on fluctuations the picture is different (Fig. 2), which suggests that further developments are needed.

In Fig. 2a we compare the experimental isothermal compressibility, which was recently measured down to -46 °C (227 K)[8], with results from a range of force-field models[9]. The dramatic increase of the compressibility in the experiment is not captured by any of the studied models. It is clear that the magnitude of fluctuations seen in experiment is severely underestimated in the simulations, though many other properties are reproduced quite accurately.

In Fig. 2b we compare the X-ray emission (XES) spectrum from experiment (top) with a computed spectrum from simulations using path-integral MD (PIMD) for nuclear quantum effects, in combination with the opt-PBE-vdW van der Waals corrected density functional[35]. XES measures the valence electronic structure and the experiment shows a clear two-peak structure in the lone-pair region (526-527 eV). The simulation, however, does not reproduce the split observed experimentally, but rather gives an average. Finally, we compare the experimental Raman spectrum of water (Fig. 2d) with that (Fig. 2c) obtained from simulations using the MB-pol force-field[36], which includes up to three-body interactions fitted to a large set of structures computed at the quantum chemical gold-standard CCSD(T) level. The overall shift to lower frequencies with decreasing temperature is reproduced, but not the bimodality observed in experiment. Some effect thus seems to be missing in the description of water that could enhance fluctuations (Fig. 2a) and cause a stricter bimodality in simulated XES (Fig. 2b) and Raman (Fig. 2c) spectra. What can be the character of such fluctuations? Clearly, since they increase with decreasing temperature, their origin and driving force must be different from normal, random thermal fluctuations in the liquid.

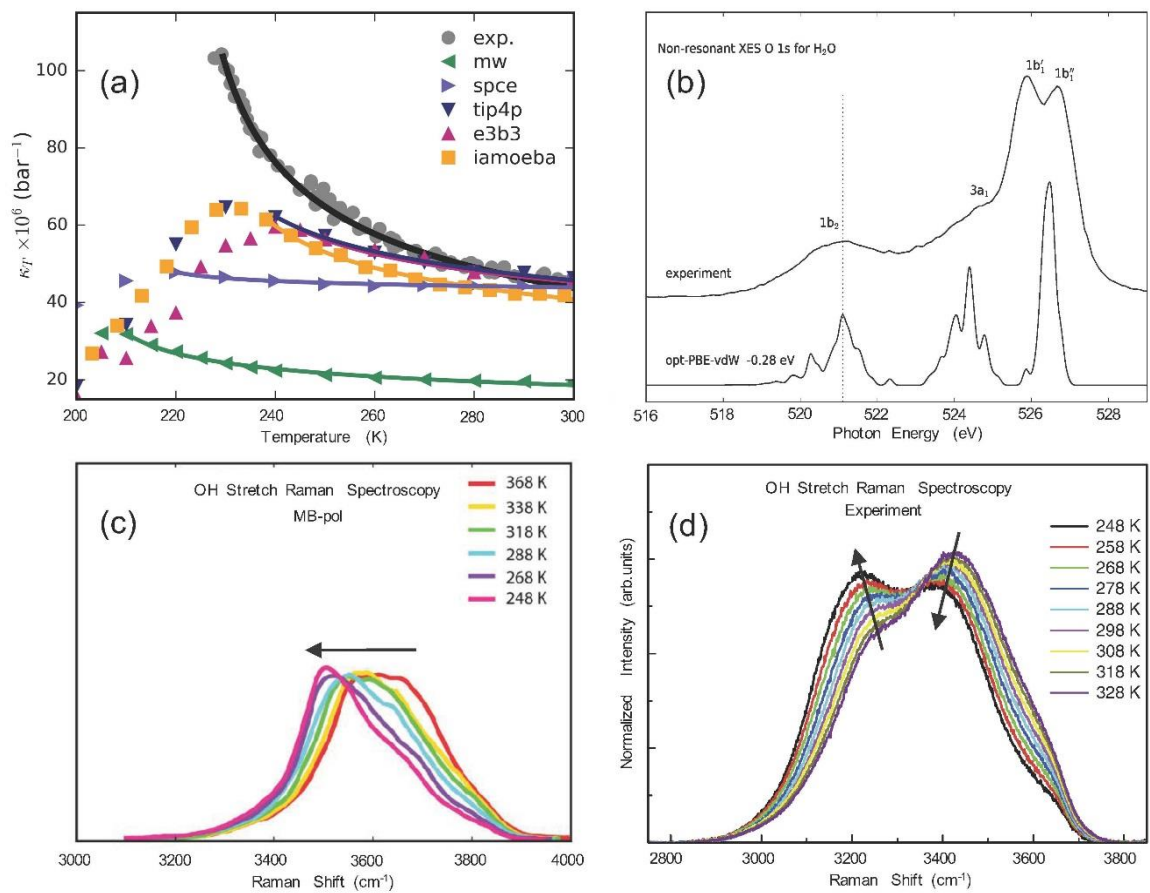

**Figure 2 (a)** Comparison of the temperature-dependent isothermal compressibility from experiment[8] (grey) and the mW, SPCE, TIP4P/2005, E3B3 and iAMOEBA simulation models. Solid lines are power-law fits to the rising part of each set of data points. Although accurate in many other respects all here tested simulation models severely underestimate structural fluctuations leading to the strong enhancement observed in experiment. (Figure adapted from ref. [9]). **(b)** Experimental XES spectrum of $H_2O$ (top) compared with computed spectra from PIMD simulations using opt-PBE-vdW (bottom)[35]. **(c)** Simulated $H_2O$ Raman spectrum using MB-pol[36] compared with **(d)** experiment[37-38]. Reprinted with permission from ref. [35].

In the following, we will present and discuss evidence from different experimental techniques and theoretical simulations, which point to fluctuations between two well-defined local environments. Before embarking on this endeavor, however, we first point out that two-state behavior has been shown to be fully consistent with thermodynamics and a prerequisite for a theory that describes the anomalies of water[17-19,39-43]. Secondly, we note that the two-state picture of water has a long history[44-49], and may be considered required to explain the properties of supercooled water[13]. For ambient water the picture of the liquid as a continuum distribution around mainly tetrahedral H-bonding is still prevalent, however, very much due to the influence of MD simulations. These typically result in a continuum picture, but severely underestimate the magnitude and specificity of the fluctuations inferred from experiment as indicated in Fig. 2.

We will begin by discussing the structure of water based on scattering techniques (X-ray and neutron diffraction and extended X-ray fine-structure (EXAFS)) and continue with information from spectroscopic (IR, Raman, X-ray absorption (XAS) and emission (XES)) techniques. We will follow up by discussing MD simulations and what aspects may need improvement in order to reproduce, in particular, structural fluctuations in the liquid. In the discussion section we will explain the origin of many of water's properties in terms of fluctuations between local HDL and LDL structures and speculate about how a deeper understanding of water can help desalination efforts.

## II. SCATTERING TECHNIQUES

X-ray (XRD) and neutron (ND) diffraction provide the most direct measurement of structure in the liquid, where the split in the first peak of the oxygen-oxygen scattering structure factor is a measure of tetrahedrality in the liquid. X-rays interact with the electronic charge density where the diffraction pattern is equivalent to the Fourier transform of the charge-density distribution in the liquid. As such, X-ray diffraction provides mainly information on the O-O and O-H correlations, while the small charge density on the hydrogens results in low information-content as regards H-H correlations. Neutrons, on the other hand, scatter against the nuclei, with highest sensitivity to the light particles (hydrogen or deuterium), such that XRD and ND provide complementary information-content on the structural arrangement. Scattering from the liquid gives the typical angularly-symmetric rings, with intensity depending on the momentum ($q$) transfer (distance from the center of the ring). This is then integrated over the $2\pi$ angle to obtain the structure factor as function of momentum transfer, which can subsequently be directly transformed to a real-space correlation function, provided that a sufficient interval in $q$ has been measured.

Since XRD and ND, measured on a single sample, give insufficient data to disentangle the three correlations (O-O, O-H and H-H), one typically adds a sample containing a different isotope with different neutron scattering length[50]. In order to avoid uncertainties due to somewhat different structure for $H_2O$ and $D_2O$[51-53], as well as the inelastic corrections due to the lighter hydrogen isotope, the contrast between $^{17}O$ and $^{18}O$ can be used[54]. Here we will focus on to what extent the H-bonding arrangement in the liquid is determined by the combination of XRD and ND.

The two dominating techniques to extract structural data, beyond the three individual pair-correlation functions from the XRD and ND scattering patterns, are empirical potential structure refinement (EPSR)[55] and reverse Monte Carlo (RMC)[56-57]. The goal for both is to develop structural models that are consistent with the scattering data, and from which further information can be obtained. In EPSR, an initial potential from some force-field description of water is used in Monte Carlo simulations of the diffraction data. The potential is then successively perturbed, based on the difference between simulated and measured data, to give agreement with the experiment[58-59]. In this approach, one thus purposely relies on a model force-field to constrain the simulation to physically acceptable solutions. This then relies on the data to correct any unintentional bias that the force-field may have introduced. Drawbacks of this approach are that, in situations where many different structural solutions are possible, still only one final potential corresponding to a specific solution is obtained, and also that the data may be insufficient to correct possible biases inadvertently introduced by the force-field.

The RMC approach, on the other hand, performs Monte Carlo atomistic moves to generate a distribution of solutions, which are maximally disordered within the constraints imposed by the experimental data, and by possible additional constraints set by the simulator[56-57]. As an illustrative example of limitations on the information content in the scattering data, we mention that with the RMC approach, unless the simulation is constrained to molecular $H_2O$ (or $D_2O$), the stoichiometry is obeyed only on average, *i.e.* one finds a distribution of $H_xO$ entities where only the average of $x$ is two. The approach is thus completely driven by the data and one may use constraints to drive the solutions in different directions and thus explore what bounds the data actually set on possible solutions. By adding more and more experimental data from different techniques that are sensitive to different structural aspects, these bounds can then be made more and more rigorous.

In the 2004 paper by Wernet *et al.*[60], XAS and X-ray Raman scattering (XRS) data were interpreted as a dominant fraction of the molecules in the liquid being very asymmetrically H-bonded, with only one well-defined donated and accepted H-bond, consistent with molecules in chains or rings. This was in contrast to the distorted, mainly tetrahedral picture of the liquid that was accepted based on earlier analyses of, *e.g.*, scattering data[61-62], and has caused significant debate[63-70].

Leetmaa *et al.*[71-72] used RMC on ND data for five different isotope compositions together with XRD to investigate whether a large number of broken H-bonds would be contradicted by the data. To this end, an H-bond criterion was added and the program was set to fit the experimental data combined with maximizing the number of intact donated H-bonds. This resulted in 74% double H-bond donors (DD) and 21% single donors (SD). Maximizing instead the number of broken H-bonds gave 81% SD and 18% DD, while still reproducing the experimental data, and losing only 0.7–1.8 kJ/mole interaction energy as measured using a range of force-fields[71]. Diffraction data thus allow a tetrahedral model, but does not exclude models with a large fraction (81%) with one broken donated H-bond as proposed by Wernet *et al.*[60]. In terms of H-bond-connectivity in the liquid, diffraction data thus allow a large range, and complementary experimental data must be applied to narrow this range.

*Extended X-ray Absorption Fine-Structure (EXAFS)*

In 2007, Bergmann and coworkers[73-74] reported an O-O pair-correlation function based on EXAFS, which was markedly different from that from XRD with a significantly sharper and higher first peak at shorter distance than obtained from XRD. In EXAFS an X-ray photon ejects an inner-shell (O $1s$) electron as a spherical wave, which is back-scattered against atoms in the environment. This gives oscillations in the scattering cross-section that depend on interatomic distance and number of scatterers; the process can be likened to a particle in a box, for which specific resonances occur when multiples of its half-wave-length coincide with the box dimensions. Could the EXAFS and XRD data be reconciled in terms of a single structural solution?

To answer this question, theoretical EXAFS signals must be computed, including multiple scattering to account for the presence of the positive hydrogens between two oxygens in an H-bond. Such calculations take too long for a standard RMC procedure based on hundreds of millions of atomistic moves, where the properties in question are recomputed after every move. To overcome this difficulty, the SpecSwap-RMC procedure[75] was developed and implemented[76]. This builds on precomputing the data for a large variety of structures that are then used to reproduce the experimental data in an RMC fashion; instead of atomistic moves, the exchanges are made among library structures based on their contributions to the fit to the experimental data. Applying this to a simultaneous fit of EXAFS signal and XRD O-O pair-correlation function resolved the discrepancy, and structural solutions that reproduced both EXAFS and XRD could be obtained[77].

EXAFS is particularly sensitive to short interatomic distances, which enhance the back-scattering signal, and furthermore to the presence of a positively-charged hydrogen between the two oxygens. The positive charge acts as a lens that focuses the outgoing spherical wave onto the accepting oxygen, as well as back onto the oxygen from which the electron was emitted. This results in a strong enhancement of the EXAFS signal, which becomes additionally enhanced if the environment is symmetric with all neighboring oxygens at similar (short) distance. EXAFS is thus particularly sensitive to a subclass of molecules in very tetrahedral environment with short and well-defined H-bonds, while XRD sees all situations with similar probability. Including a subset of molecules in very well-defined, tetrahedral local environment, with the remainder in more disordered structures, thus resolved the seeming discrepancy and emphasized the importance of using complementary experimental information when deducing structure in the H-bonding network[77].

*Wide-Angle X-ray Scattering (WAXS)*

By using very high-energy X-ray photons it is possible to extend the measured $q$-range in XRD to high-enough values that a direct Fourier transform can be applied to extract the O-O correlation function with very small uncertainty[78-79]. The O-H contribution must still be removed, but once this is done, a high-accuracy O-O pair-correlation function for ambient water[79] and, in particular, the temperature-dependence of the various peaks, can be reliably determined[78,80]. Most notably, it was found that the second correlation (at ~4.5 Å) becomes well-established only below the compressibility minimum at 46 °C[78] and that this is accompanied by the growth of specific correlations at 8.5 and 11 Å, as well as weaker radial correlations out to ~17 Å[80]. The peak at 4.5 Å is associated with tetrahedral H-bonding and the association with growth of the additional peaks at longer distance indicates the appearance and growth of collective fluctuations into tetrahedral structures, which can be associated with the increased compressibility below 46 °C. As further indication of a change in behavior around the compressibility minimum, there are changes in slope with temperature for properties as varied as, *e.g.*, the thermal conductivity (64 °C), the refractive index (60 °C), and the conductance (53 °C) as compiled by Maestro *et al.*[81].

*Small-Angle X-ray Scattering (SAXS)*

Extending the $q$-range instead to lower values towards zero $q$-transfer in small-angle X-ray scattering (SAXS) gives information on larger-scale fluctuations through the inverse relation between momentum space and real space. This was used by Huang *et al.*[20,82] who estimated the typical size of tetrahedral fluctuations in ambient water to be around 10 Å. The suggested local heterogeneity in terms of tetrahedral fluctuations in a predominantly high-density environment was criticized[63-64,70,83-84], but is in good agreement with the observed temperature-dependent correlations in the O-O pair-correlation function[80] discussed above, as well as with the spectroscopic evidence to be discussed below. The conclusion is furthermore supported by MD simulations[85-86] using the TIP4P/2005 force-field[87], although it should be noted that the enhancement of the low-$q$ scattering upon deep supercooling is severely underestimated by the simulations[86].

Extending the low-$q$ scattering measurements of the structure factor, $S(q)$, to zero $q$-transfer at different temperatures gives information on the temperature-dependent isothermal compressibility, $\kappa_T$, through the thermodynamic relationship $S(0) = k_B T n \kappa_T$, where $k_B$ is the Boltzmann constant, T is the temperature (in Kelvin), and $n$ the number density. Since $\kappa_T$ is related to density fluctuations, the underestimated enhancement of $S(0)$ in the simulations at low temperatures[86] indicates that structural fluctuations at low temperatures are underestimated in the simulations.

The connection to the isothermal compressibility provides an entrance to exploit SAXS to determine $\kappa_T$ to lower temperatures than previously possible. This was recently achieved by performing SAXS measurements on micron-sized water droplets that are cooled ultrafast in vacuum and measured using femtosecond coherent pulses from an X-ray free-electron laser (XFEL)[8-9]. These measurements are extremely challenging considering that the kilometers-long, linear electron-accelerator generates "bullets" of X-rays with diameter ~3 microns (3·10$^{-6}$ meters) which have to hit water droplets of ~10 microns while they fall in vacuum. However, the technical difficulties could be overcome and the SAXS signal measured on liquid water down to -46 °C (227 K)[8-9]. At these low temperatures, significantly below the previously established temperature of homogeneous ice nucleation, the water droplets crystallize on a sub-millisecond time-scale, which necessitates the ultra-fast probing that has become possible with the availability of XFELs.

As result of these measurements, it could be shown that the isothermal compressibility exhibits a maximum at -44 °C (229 K)[8], rather than the divergence that had been proposed based on extrapolation from higher temperatures[7]. A divergence is expected upon approaching a critical point, while a maximum is expected along the continuation of the coexistence line into the one-phase region beyond a critical point, the so-called Widom line. This line in the phase diagram separates regions dominated by either species (HDL or LDL) and along this line fluctuations are maximal as it corresponds to a 50:50 average composition in terms of the fluctuating species[88]. The magnitude of the maximum in $\kappa_T$ depends on the distance to the critical point and, by comparison with simulations using the iAMOEBA model[89], the LLCP in water was proposed to lie at ~800 bar pressure[8].

To summarize the observations from scattering measurements, the pure HDL and LDL liquids and transformation between them have been identified in the ultraviscous regime using a combination of WAXS and XPCS upon heating HDA at ambient pressure[28]. A maximum has been observed in the temperature dependence of $\kappa_T$, consistent with a Widom line that indicates the presence of an LLCP at higher pressure[8]. The proposed LLCP has thus been approached both from below the crystallization temperature by starting from amorphous ice and from above the temperature of homogeneous ice nucleation by using micron-sized water droplets.

Using XRD with a very high *q*-range, a change towards more well-developed tetrahedral fluctuations below the minimum in $\kappa_T$ was found[78], which could furthermore be connected to the growth of well-defined correlations out to ~11 Å[80], that indicate the appearance of collective fluctuations[90]. This is consistent with earlier determinations of the spatial magnitude of density heterogeneities due to tetrahedral fluctuations using SAXS[20-21]. We note finally, that the combination of XRD and EXAFS, as well as XES (see below), requires a minority subset of structures in the ambient liquid, with very well-defined tetrahedrality and short H-bonds that connect the participating molecules[77].

## III. SPECTROSCOPIC TECHNIQUES

*Raman spectroscopy*

The temperature-dependent Raman spectrum of liquid neat water is bimodal with a transfer of intensity from the high-energy side to that at low energy upon cooling (Fig. 2d)[37-38,91-95]. Scherer *et al.*[91] used the polarization dependence to show that molecules contributing to the low-energy side of the spectrum are in a highly symmetrical, tetrahedral environment with two donated and two accepted hydrogen bonds, while the local environment of molecules contributing to the high-energy peak is highly disordered. This is consistent with the expected temperature-dependence of H-bond formation, with more disorder and broken or distorted H-bonds at higher temperature. It is also consistent with the effects on the spectra from applying pressure[95], which inhibits fluctuations into low-density tetrahedral structures and reduces the intensity on the low-energy side of the Raman spectrum. Walrafen[93] showed furthermore that, upon heating the liquid, the integrated intensity-gain above the isosbestic point (point where the intensity is independent of temperature, see Fig. 2d) at 3425 cm$^{-1}$ in the OH-stretch vibration very closely matches the intensity-loss integrated below this point, *i.e.* fully consistent with temperature-dependent conversion between two specific types of H-bonded species.

*Infrared Spectroscopy*

Further compelling evidence for interconversion between two specific local structures in the liquid is provided by Maréchal[96], who showed that the temperature-dependent (0 – 80 °C)

IR spectrum (0 – 4000 cm$^{-1}$) can be decomposed as a temperature-dependent linear combination of two temperature-*independent*, complete spectra. Since the IR spectrum is sensitive to H-bond structure, this provides a strong indication of the existence of two rather specific classes of H-bond situations in the liquid, tetrahedral (LDL) and disordered (HDL), consistent with the analysis based on polarized Raman OH-stretch data[91] and the XRD, EXAFS and XES data above.

*X-ray Spectroscopies*

X-ray spectroscopies in absorption (XAS) and emission (XES) measure the electronic structure locally around the probed molecule in the liquid or solid[22,97-100]. In absorption, an X-ray photon is used to excite an O 1$s$ core-electron into empty valence and continuum states. XES is connected to XAS as the decay of a valence electron from an occupied level to fill the 1$s$ core-hole, with the released energy carried away by an X-ray photon. Since the 1$s$ orbital is strongly localized, and an overlap between the initial and final levels is required for a transition, both spectroscopies provide very local probes of the electronic structure – unoccupied states in XAS and occupied in XES - which depends on the H-bonding situation[35,101]. Both spectroscopies obey the dipole selection-rule, which makes them sensitive only to the local $p$-character of the unoccupied (XAS) and occupied (XES) states due to the spherical symmetry of the 1$s$ core level. XAS is thus particularly sensitive to donated H-bonding due to the $p$-character in the internal O-H antibonding states, which are located mainly on the hydrogens, while in XES the highest intensity comes from the doubly-occupied 2$p$ lone-pairs.

From the traditional picture of liquid water as mainly tetrahedrally H-bonded, an XAS spectrum similar to that of ice would be expected, but broadened and smeared out by disorder since the excitation is into antibonding states whose energy position depends on the bond distance. Instead, the dominant post-edge at 540 eV of ice is all but lost and a main-edge at 537 eV and sharp pre-edge at 535 eV become prominent[60,98-99,102-103]. This was interpreted by Wernet *et al*.[60] as most molecules in the liquid being in an asymmetric H-bonding situation with only one well-defined donated and accepted H-bond, indicating chains or larger rings in the liquid[60].

*XES*

The core-hole-state generated by XAS is very highly excited (~540 eV) and decays on a time-scale of the order 4 femtoseconds[104]. When this decay occurs through emission of an X-ray photon, it may be detected in XES and give a direct measurement of the atomic 2$p$ character in the occupied orbitals on the probed molecule. We focus on the non-bonding 1$b_1$ lone-pair, which does not get vibrationally excited and remains sharp[20,105-111]. Hydrogen-bonding in crystalline ice shifts the peak to ~1.5 eV lower energy compared to gas phase. Interestingly, liquid water shows *two* sharp lone-pair peaks that interconvert, but do not broaden with increasing temperature (Fig. 2b); the peak at low emission energy (close to the position in ice) goes down in intensity, but remains in fixed position at higher temperature, while the peak at high energy increases in intensity and disperses towards the gas-phase position[20,23,106,112]. This is consistent with the existence of two specific types of local structures, very tetrahedral (LDL) and very disordered (HDL), in the liquid[20,22,106-109]. However, removing a screening 1$s$ electron effectively converts the oxygen to a fluorine (Z+1 approximation[113-114]) which generates strong vibrational excitations through core-hole-induced dynamics, that can even lead to dissociation[115]. An alternative interpretation of the peaks has been proposed as due to, respectively, intact and dissociated molecules[105,111,116-118]. This interpretation has recently, through direct measurement of the contribution from dissociation[119], been shown to be inconsistent, which favors the picture of interconversion between two specific local structures in the liquid, as also concluded from the IR and Raman studies above.

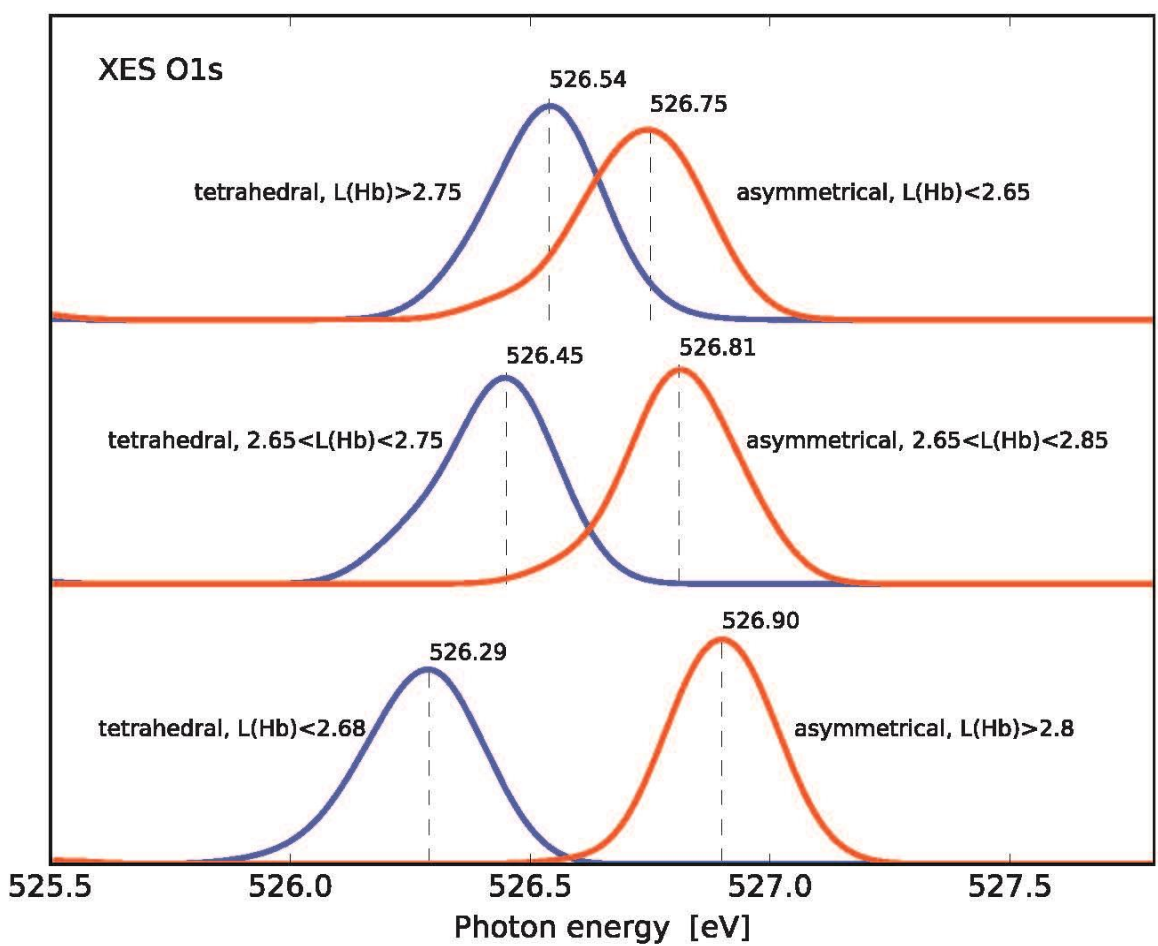

**Figure 3.** Computed lone-pair XES peak positions as function of H-bond symmetry (tetrahedral and asymmetrical) and H-bond distances. The split increases toward the experimentally observed by considering a subclass of very tetrahedral molecules with short H-bonds and asymmetrically H-bonded with long H-bonds. Reprinted with permission from ref. [35].

As seen from Fig. 2b, typical MD simulations result in a single peak at the lone-pair position in XES when computing the contribution from all molecules in a snapshot of the simulation[35]. Experiment, however, shows a rather sharp double-peak feature with energy separation and relative intensity depending on the temperature[20,100,105-109,112,120]. The peak position in the simulations is intermediate between the two peaks observed experimentally, which indicates that the instantaneous structure in the liquid is somehow intermediate between two more well-defined structures in the real liquid[35]. This is similar to the simulated Raman spectrum in Fig. 2c, which exhibits the trend towards lower energy with decreasing temperature, but not the clear bimodality and intensity transfer between two peaks that is observed experimentally (Fig. 2d).

The lone-pair peak position in XES, computed for local structures in the simulated liquid depends on the local H-bonding, where fully H-bonded molecules in tetrahedral environment were found to contribute at lower emission energy than molecules in asymmetrical environments with some broken or weakened H-bonds[35]. However, applying this criterion by itself when computing the spectrum resulted in a significantly too small computed split between the peaks in comparison with the experiment (Fig. 3, top). The computed peaks are very broad, however, and, by restricting H-bond distances within each class (tetrahedral and asymmetrical), it was found that the computed split is rather sensitive to this parameter (Fig. 3, middle and bottom): in order to obtain a split comparable to experiment it was found necessary to restrict the class of tetrahedral species to molecules with very short (<2.68 Å) and well-defined H-bonds. The asymmetrical species, on the other hand, were found to require long (>2.8 Å) H-bonds, as also expected for an HDL local structure which squeezes in a fifth neighbor between the first and second coordination shells[35]. We thus have a consistent interpretation of XRD, EXAFS and XES in terms of a subset of molecules in very well-defined tetrahedral LDL-like environment, while the majority at ambient conditions prefer a more HDL-like environment, with expanded first coordination shell to accommodate a fifth neighbor and some broken or weakened H-bonds. MD simulations do contain such structures, but they constitute a minority with most molecules in local structures that constitute more an average between these situations. In Section IV we will discuss requirements on simulations to remedy this situation.

*Connecting XAS and XES and Raman Spectroscopy*

By selecting different energies for creating the core-hole in XAS, the absorption features in XAS can be connected to emission features in XES[20]. By exciting at the post-edge in XAS, which contains significant contributions from tetrahedrally H-bonded molecules, the emission peak assigned to tetrahedral species is enhanced[20]. Excitation at the sharp pre-edge, which grows in intensity as the liquid is heated and has been assigned to weakly coordinated OH-groups, results in only the peak due to disordered species appearing in XES, consistent with only these species contributing to the sharp pre-edge. By tuning the XAS excitation energy it is thus possible to select specifically-coordinated species in the liquid for study[20].

XES is thus connected to XAS through energy-selective excitation, but there is more information to be obtained by connecting also to vibrational spectroscopy. XES is an energy-loss spectroscopy, in which a photon is scattered from the system, losing part of its energy; this is similar to optical Raman spectroscopy. By resolving losses due to vibrational excitations in XES, it is possible to directly measure the OH-stretch energy for the different species selected through the XAS process[111,121-122].

The energy of the OH-stretch vibration of the proposed weakly H-bonded OH-group of the disordered species (HDL), contributing to the pre-edge, is found to be 0.45 eV, which is close to the gas-phase symmetric (0.453 eV) and asymmetric (0.465 eV) stretch energies, thus confirming the assignment of the pre-edge as due to very weakly H-bonded OH-groups[122]. By scanning over the XAS spectrum and monitoring the vibrational loss features, a direct connection between species selected in XAS and their contribution to the Raman spectrum could be established, where, *e.g.*, the post-edge with mainly tetrahedrally coordinated molecules contributes to the low-energy part of the Raman spectrum, while pre-edge excitation contributes on the high-energy side[121]. There is thus a full consistency and direct experimental connection between the X-ray spectroscopies (XAS and XES) and IR and Raman vibrational spectroscopies for water.

To summarize the spectroscopic information, we note that both vibrational Raman and electronic XES exhibit clear bimodality in terms of two peaks and intensity transfer from one to the other with changes in temperature. The peak on the low-energy side of the Raman spectrum (left in Fig. 2d) has been shown experimentally, using symmetry-resolving techniques[91], to be due to molecules in tetrahedral local environment, which is also consistent with this peak growing in intensity as the liquid is cooled. In the same study, the peak at higher energy (right in Fig. 2d) was shown to be due to molecules in asymmetric H-bonding environment. This is fully consistent with the assignment of XAS and XES, as demonstrated through selective probing of different local H-bonding structures in the liquid and connecting to the Raman spectrum[121-122]. The observed bimodality in terms of spectral features indicates that the molecules in the liquid spend more time in the corresponding H-bonded environments than in structures intermediate between them. This is further underlined by the decomposition of the IR spectrum in terms of a temperature-dependent linear combination of two temperature-independent complete spectra.

## IV. MD SIMULATIONS

In Fig. 4 we illustrate different possible situations that are consistent with the experimental data discussed above, and which can be valid for water at different combinations of temperature and pressure. Going from the bottom to the top in the figure corresponds to a combination of going from high to low pressure and from low to high temperature. At the bottom (Fig. 4a) we have the two different pure forms of liquid water, for which the transformation from HDL to LDL has recently been observed (LDL in blue and HDL in yellow)[28]. Beyond the proposed LLCP there are large fluctuations into the minority species, which is LDL on the HDL-dominated side (right) and HDL on the LDL-dominated side (left) of the Widom line (dashed in the figure). At the Widom line the distribution is 50:50 on average[88]. Based on the bimodality observed in IR, Raman and XES, structures representing a structural average in between the two extremes, indicated with fuzzy green, constitute a small fraction. The HDL and LDL structural species rearrange into each other through fluctuations on a time-scale longer than the H-bond lifetime. The farther away from the LLCP one moves (Fig. 4b-d) the wider the fluctuating region becomes, while the length-scale of the fluctuations diminishes. It should be noted that, the high-precision temperature-dependent XRD data[78,80], as well as the SAXS data[20,82], indicate that the cross-over from the situation in Fig. 4d (right) to that in Fig. 4c (right) occurs around 50 °C, *i.e.* collective structural fluctuations appear at temperatures and pressures that are relevant to life. Fig. 4e, finally, illustrates the simple liquid situation, where the liquid only encompasses a homogeneous single structural environment with normal thermal fluctuations. These involve only very local structural variations reminiscent of LDL and HDL, while most of the molecules are in H-bonded structures that are intermediate between the two extremes.

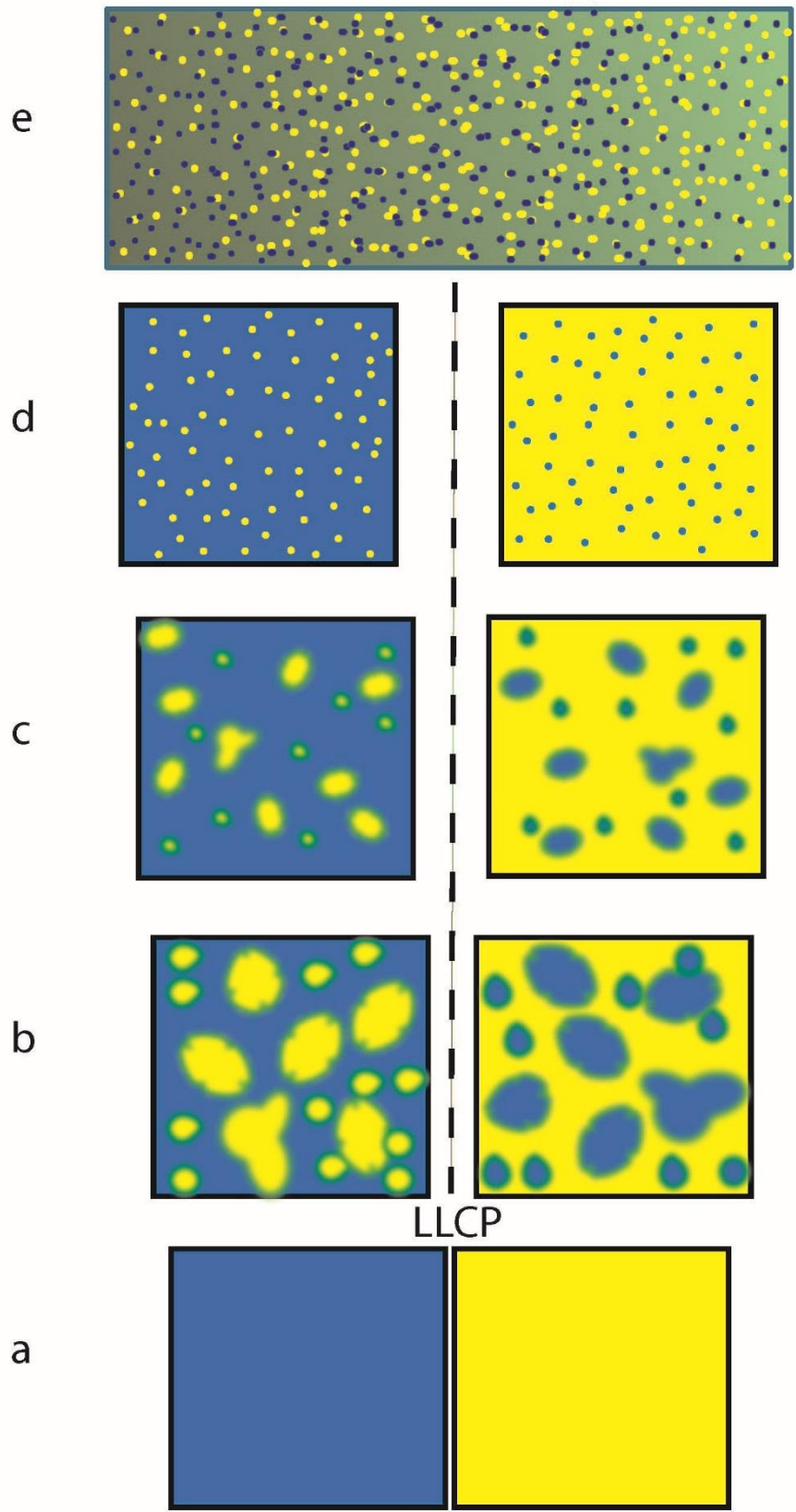

**Figure 4.** Illustration of different possible scenarios for fluctuations in liquid water at different temperatures and pressures. Blue illustrates LDL and yellow HDL. **(a)** The pure liquid forms as observed using WAXS and XPCS in the ultraviscous regime[28] with the coexistence line separating them. Beyond the proposed LLCP there are **(b)**

large fluctuations in the one-phase region where on one side of the Widom line (dashed) LDL dominates and on the other HDL. The fuzzy green borders indicate the transition region between the two local forms. **(c)** Moving farther from the LLCP, the one-phase region becomes broader while fluctuations diminish and **(d, e)** sufficiently far away, the fluctuations become small enough to represent an ideal mixture.

There are two simple pictures that can account for the observed bimodality in the dynamics of the fluctuations in the heterogeneous case (Fig. 4b-c). One is that the fluctuations behave similar to the motion of a classical pendulum moving between two turning points: the velocity is zero at the turning points and at a maximum at the point in the middle. In this picture the molecules of the liquid will spend most of the time at the extreme points and little time in between as they convert. This leads to an extreme heterogeneous situation, where two well-defined structural classes develop on a temperature-dependent length and time scale. The interfacial region between the end-points (local HDL/LDL) would contain few molecules, since only a short time is spent in this region. Another hypothesis may be that the LDL structural fluctuations appear as attempted, but frustrated, growth around specific, more long-lived templates that begin to form below ~ +50 °C. Here a number of authors have pointed out the favorable stability of clathrate structures, *e.g.*, references [45-46,95,123].

The similarity in frequency of the OH-stretch maximum in Raman measurements of highly supercooled liquid water to that in clathrates was noted by Walrafen[95] and Pauling proposed that many water properties could be understood based on clathrate structures in the liquid[45]. X-ray diffraction of supercooled water has also been interpreted as indicating the presence of clathrate structures[124], although, as shown by the RMC modeling discussed in Section II above[71-72], XRD allows a multitude of structural solutions.

Clathrates are built from open, ball-like structures, largely based on pentagonal rings, which leads to favorable angular correlations[123], but a low density, which has been taken as argument against such structures in the liquid. However, considering both HDL and LDL as simple liquids[23] that each has increasing density with decreasing temperature, the density maximum is easily understood as the point where low-density tetrahedral fluctuations overcome the density increase of the HDL component that dominates at higher temperatures. Similarly, there should then also be a density minimum when the liquid has been converted to LDL, for which the density also increases upon further cooling. This density minimum has indeed been observed for water in confinement[125-126] and in MD simulations of deeply supercooled water, *e.g.*, reference 86. The density of an LDL component should thus be closer to that at the density minimum than to that of ambient water. Fluctuations involving clathrate-like components can thus not be excluded based on density arguments. They have been proposed based on careful analysis of experimental Raman spectra[95], but have not been directly structurally identified in the liquid.

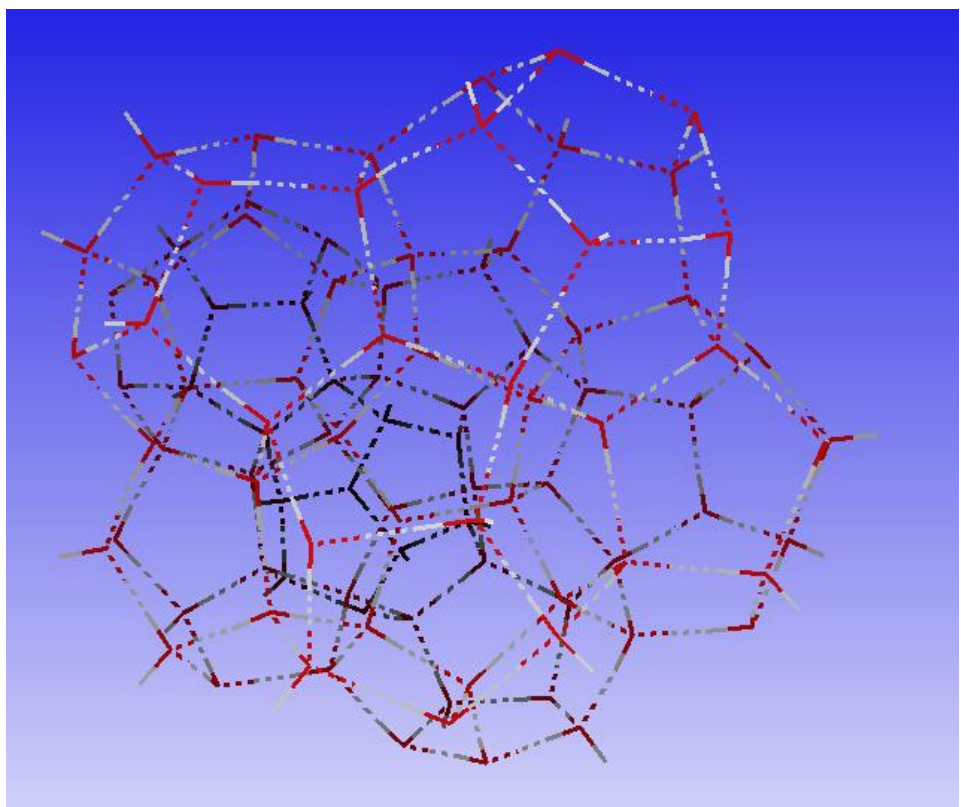

**Figure 5.** Illustration of a clathrate-like structure built from fused dodecahedra (20 molecules each). Dashed lines indicate H-bonds between molecules which are located at the vertices of the connecting lines. The internal OH bond in each molecule is indicated by solid sticks colored red-grey.

A hypothetical clathrate-like structure, built from fused dodecahedra, is shown in Fig. 5 as an illustration. It is clear from the figure that, if such structures appear, the core region could be expected to be stabilized by favorable H-bonding. It is also clear that the growth of such hypothetical structures would be hindered through the increased angular strain in outer layers, which could give rise to a narrow surrounding region of intermediate H-bonded structures as molecules from the HDL environment attempt to attach to the template, but become unstable and return to HDL. Clathrate-like structures have not been reported in MD simulations of ambient or supercooled water, which, however, does not exclude that such structures could still be relevant for real water.

Water does represent a severe challenge to theoretical simulations, not only due to the delicate balance between different counteracting interactions of similar magnitude, but also due to the emergent character of the resulting properties, which requires a large number of molecules to develop. It is clear that highly accurate computational approaches are needed to reliably capture this balance and significant progress has been made in terms of many properties, such as the RDFs. Recent developments with parameterization of up to three-body interactions at the CCSD(T) level of quantum chemical accuracy hold promise[30], but the resulting structure is still unimodal, both in terms of computed XES[35] and Raman spectra (Fig. 2c)[36]. In another approach, VandeVondele and coworkers[31] have exploited algorithmic and computer developments to perform fully *ab initio* periodic MP2 quantum chemical Monte Carlo simulations of 64 water molecules, but considering the experimental observation of an eighth *radial* coordination shell at ~17 Å, significantly larger simulation boxes will be necessary, which will be a challenge to this approach. Simulations based on DFT[127-128] scale better with size of the system, but a simulation box size of 30-40 Å still represents a severe challenge, although impressive progress is being made[32,129-131]. Nuclear quantum effects (NQE) are furthermore important[132-134] and have been proposed to be decisive in determining the difference in entropic contribution from HDL and LDL[21]. The expense of including NQE has been significantly reduced by developments in terms of colored noise Langevin thermostats[135] and quantum ring-polymer contraction[136].

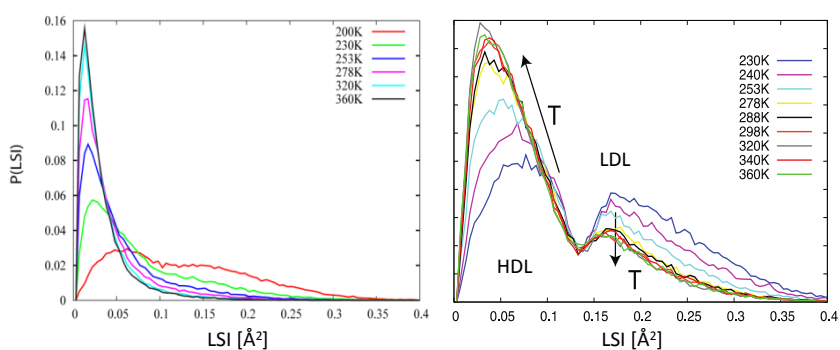

**Figure 6.** LSI distributions obtained from (left) the real structure and (right) energy-minimized "inherent structures" in simulations of TIP4P/2005 water as function of temperature. An isosbestic point is seen around LSI = 0.13-0.14 Å$^2$ and is present at all temperatures in the inherent structure, but only as a weak minimum at the lowest temperature in the real structure.

In Fig. 6 the local structure index[137], LSI, is used as order parameter to demonstrate that, even in simulations with classical force-fields, the *inherent structure* (Fig. 6, right) is perfectly bimodal[88]; the inherent structure is obtained by quenching the instantaneous structure into the nearest local minimum on the underlying potential energy surface, thus removing thermal motion [138]. The 3N+1 potential energy landscape (PEL) that is thus revealed (N being the number of atoms in the simulation) has been analyzed in connection with supercooled liquids and the glass transition[139], as well as connected to the anomalous properties of water and a liquid-liquid transition or a first-order-like transition between the amorphous ices LDA and HDA[140-142]. As seen from Fig. 6, the LSI order parameter reveals a clear bimodal distribution in terms of two basins corresponding to HDL and LDL, similar to the two basins discussed in Ref. 141 in connection either with HDL and LDL or HDA and LDA.

In the real structure (Fig. 6, left), including temperature, the bimodality is, however, lost except at the lowest simulated temperature (200 K). The LSI reflects the degree of order in the first and second coordination shells and gives a high value for tetrahedrally strongly ordered local coordination (LDL) and a low value for close-packed, disordered local coordination (HDL).

The bimodality in the inherent structure implies that the multidimensional potential energy surface, on which the simulation evolves, contains two qualitatively different types of projections relating to *local* configurations of molecules in agreement with XAS [60,98-99] and XES [106,112] experimental data, as well as with IR[96] and Raman[38,91-92,94]. We note that the populations at ambient conditions in the inherent structure, ~25% high-LSI (LDL-like) and 75% low-LSI (HDL-like)[88], reflect closely the experimental estimates[20,84,106] of the two components in XES and XAS/XRS[60,98], as well as those of Taschin *et al.*[143] based on the optical Kerr effect and Xu *et al.*[144] from IR data on confined water. We observe further, that the behavior of the two peaks in the LSI distribution matches what is observed from XES, where the lone-pair feature, assigned as due to tetrahedral species (LDL-like), only loses intensity with increasing temperature, but shows no dispersion; this is also seen for the high-LSI distribution in Fig. 6 (right). The other peak in XES, close to the gas phase position and assigned as due to disordered species (HDL-like) gains intensity and disperses toward the gas phase peak position with increasing temperature; this is closely matched by the low-LSI distribution in Fig. 6 (right).

We thus have a direct correspondence between the two LSI components and the spectroscopic observations. However, including temperature removes the bimodality. The simulation exhibits the prerequisites for generating a bimodal spectroscopic signal in that the underlying potential energy surface on which it evolves does contain two minima, consistent with the interpretation of the spectra. The minima are, however, seemingly too shallow such that, when a finite temperature is included, an average structure is obtained. In this picture the two minima would thus need to be deeper to constrain the simulation more closely to either minimum.

An alternative is to remove thermal energy by simulating deeply supercooled and pressurized water. In this case, many different water models have indicated the existence of an LLCP associated with a liquid-liquid transition[12-13,15-16,24,29,39,42,145-156]. However, since the LLCP in the models is located at very low temperature, these simulations are challenging and require very long equilibration times. Since ice is significantly more stable in this metastable region, the interpretation in terms of a liquid-liquid transition in the ST2 force-field model of water[16] was challenged by Chandler and Limmer in a series of papers[157-159] and claimed to instead be a liquid-solid transition. Debenedetti and coworkers[12,151] used the same model and, in contrast, found unequivocal evidence of a reversible transition between two basins corresponding to LDL and HDL in the free-energy landscape, as well as locating the transition to ice. Since they used the same model as Chandler and Limmer, but with different results, this led to a heated exchange[151,160-161], which was only resolved when it was shown that the Chandler and Limmer simulation code had conceptual problems[146,162]. Thus, it is now clear that the ST2 model of water, as well as many other water models, supports a liquid-liquid transition but, when simulating *ambient* water, the experimentally observed bimodality and structural fluctuations, that would be expected from the existence of an LLCP, become washed out.

Fig. 7 illustrates in a simplified picture what effects need to be considered and improved in simulations in order to better represent the bimodality in structural fluctuations that the experimental data above suggest also for ambient liquid water. Since we discuss bimodality, we will illustrate the result of adding specific interactions through a one-dimensional free-energy surface condensed to two basins representing LDL and HDL at ambient conditions; the 9N (N being the number of molecules) degrees of freedom have thus been condensed into one, qualitatively illustrating the free-energy balance between LDL- and HDL-like environments.

Fig. 7a illustrates the qualitative behavior in a classical force-field simulation, where the minima are present, but not deep enough to constrain the simulation to either of the two wells; the resulting distribution (red) becomes an average due to the available thermal energy (dashed).

Electronic structure cooperativity in H-bond formation, which strengthens individual H-bonds if a more extended tetrahedral network is created [163-164], is not included in common classical non-polarizable force-fields. Extending the description to *ab initio* MD simulations, using DFT at the generalized gradient level, enhances the minimum corresponding to tetrahedral or LDL-like structures, as indicated in Fig. 7b. The balance is, however, shifted too much towards tetrahedral H-bonding and LDL as demonstrated by the melting temperature of ice, which is found to be around 420 K when simulated using the PBE or BLYP functionals[165].

The non-directional van der Waals interaction can offset the too strong directional H-bonding as illustrated in Fig. 7c. HDL is a more close-packed structure, which is favored by more non-directional interactions. Including the van der Waals interaction thus enhances the other, HDL-like, minimum in the simulations, as shown by the melting point of BLYP ice being lowered by 60 degrees to 360 K when van der Waals interactions are included[166]. MD simulations using functionals including van der Waals interactions fully *ab initio* [32,167-169] have reported that the very overstructured, LDL-like O-O RDF obtained with GGA functionals becomes a very feature-less, HDL-like RDF when instead applying the vdW-DF2 or optPBE-vdW non-local correlation functionals [168]; the recent work by Del Ben *et al.*[32], investigating a range of techniques to include non-local correlation, underlines the importance of a proper balance between H-bonding and non-directional van der Waals interactions.

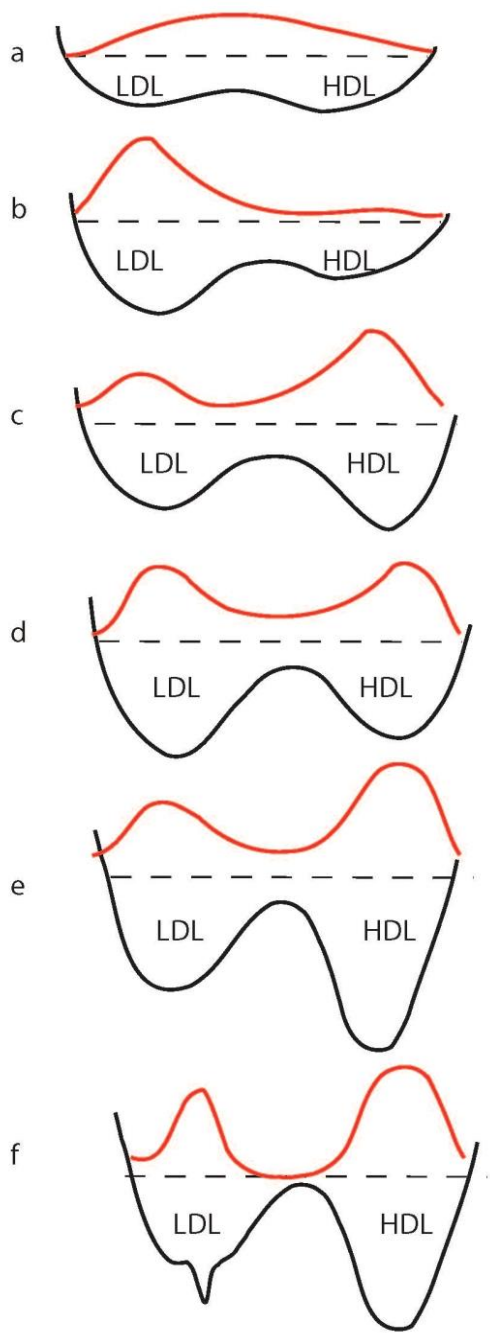

**Figure 7**. Illustration of the effect of various contributions on the balance between HDL and LDL. (**a**) Classical force-field. (**b**) Add electronic cooperativity through DFT. (**c**) Add van der Waals interaction. (**d**) Increase the box size. (**e**) Add nuclear quantum effects. (**f**) Allowing enough time for the simulation to find preferred templates for tetrahedral LDL fluctuations.

The size of the simulation box is of particular importance for the LDL-like fluctuations, as illustrated in Fig. 7d. The analysis of the temperature-dependent extended O-O RDF's by Schlesinger *et al.*[80] revealed shell structure out to around 17 Å becoming evident upon cooling. Since the RDF describes the *radial* distribution, then a box size greater than ~35 Å would be needed to allow fluctuations into *one* such tetrahedral patch to fully develop; a standard 64 molecules *ab initio* MD simulation at experimental density uses a box of 12.42 Å side.

Fig. 7e illustrates the importance of nuclear quantum effects[170] that have been proposed to discriminate between LDL and HDL[21,171], since it is only when the stiffer modes in tetrahedral coordination (LDL) are fully quantized that these modes become thermally inaccessible and their contribution to the entropy eliminated. The less restricted motion in the HDL environment, with broken or weakened H-bonds, leads to softer vibrational motion and thermally accessible quantized energy levels at ambient temperatures which contribute to the entropy.

Finally, until now only up to short-range three-body interactions have been included in developing the force-fields that hold the potential to span both the spatial and temporal range necessary to fulfill the requirements outlined above. Fig. 7f illustrates the potential effect of specific clathrate-like templates as basis for fluctuations, which could generate the rather sharp bimodality observed from IR, Raman and XES spectroscopies by forming a core template for fluctuations as they attempt to grow. This is indicated by the sharp specific well and absence of significant intensity in the transition region between the HDL and LDL basins.

However, even for as simple a structure as a 20-molecules dodecahedron there are 30,026 different ways to organize the same number of H-bonds, while still obeying the ice-rules[172]; these span a significant range in energetic stability depending on the topology of the H-bond network and would be only the simplest building-block in pentagon-based, clathrate-like templates for fluctuations. Even if all the above requirements are fulfilled in terms of description of the interactions, size of simulation box and inclusion of nuclear quantum effects, it will likely require a presently unattainable computer time for such templates to appear simply through statistical motion in the simulation box. This is qualitatively illustrated by the rather sharp local minimum in the potential energy surface in Fig. 7f.

One additional, strictly quantum mechanical effect may also be of importance for obtaining the correct instantaneous structure of the liquid. The overall wave function for a system of protons must be antisymmetric with respect to exchange of the two protons since they are fermions, while for deuterons (being bosons) it must be symmetric. An antisymmetric spin wave function for the two protons thus enforces a symmetric spatial wave function for a system of protons (*para*-$H_2O$), while the opposite is true for the symmetric spin case (*ortho*-$H_2O$). The spatial wave function of the two protons relates to rotations, such that an *ortho* water molecule cannot be in the lowest rotational state. The energy difference in gas phase is small[173], 23.8 cm$^{-1}$, but in condensed phase rotations become hindered with potentially larger energy differences when the relevant modes become librations and translations. Interestingly, the spin statistics give a 1:3 *para* to *ortho* ratio similar to what is found for the LDL to HDL ratio in the inherent structure in TIP4P/2005 water[88], as well as experimentally proposed from X-ray spectroscopy[20,60].

Although significant efforts have been made to take into account wave function aspects of both electrons and light nuclei[174-180], the restrictions on the nuclear wave functions because of requirements on exchange of identical fermions have not been specifically treated. Notably, triplet-coupled (*ortho*) water molecules cannot be in the rotational ground state, which could make it unfavorable for *ortho* molecules to participate in fluctuations into tetrahedral, LDL-like environments. Similar requirements pertain for the bosonic wave functions of $D_2O$ water.

It is not clear to what extent the requirements on the proton (deuteron) wave function affect the energetics and dynamics in the liquid. Is spin conversion through hyperfine interactions and proton (deuteron) exchange rapid enough to follow the picosecond dynamics of H-bond formation and breaking or will the wave function requirements induce preferences in the conversion between local structures? For water molecules confined in $C_{60}$ fullerenes the time-constants for interconversion at room temperature was determined as 16-30 s (depending on oxygen isotope)[181] and, in a very recent experimental study[182] it was shown that *ortho*- and *para*-water exhibit different reactivity in a prototype reaction.

There are thus significant challenges remaining to simulators aiming to obtain all properties of this fundamental liquid correctly from a single, fully consistent simulation. These concern both developing accurate descriptions of the molecular interactions and a fundamental understanding of the quantum properties of the light particles in terms of proton delocalization, as well as wave function properties through nuclear spin-coupling. These must, furthermore, be possible to efficiently scale up to length-scales of tens or possibly hundreds of thousands of molecules and extended to time-scales that are sufficient for a fully ergodic sampling of the distribution of LDL fluctuations. The latter may occur based on structural templates that live on a very different time-scale than the normal H-bond dynamics in the liquid, which will further complicate the life of the simulator aiming for the "ultimate description" of water.

Having said this, a large number of properties are indeed well-described by present models[30,34], although it is clear that earlier conclusions from Raman data[91-93,95,183] and more recent IR data[96], as well as from XAS and XES data[22-23,35,97,99,106,112] and scattering[8-9,86], are not well-described and indicate possibilities for improvement of the simulation models.

## V. DISCUSSION AND CONCLUSIONS

We have presented a range of experimental data that indicate that the local structure of normal liquid water under ambient conditions is dominated by HDL-like environments with fluctuations into strongly tetrahedral environments with very well-defined and short H-bonds. The isosbestic point observed in the temperature-dependent OH-stretch Raman signal[37-38,92,94], the split between the two lone-pair peaks in XES[35], as well as the decomposition of the temperature-dependent full IR spectrum as two temperature-independent spectra[96], indicate that structures intermediate between HDL-like and LDL-like do not contribute significantly. This could be due to rapid transitions between the two endpoints, *i.e.* HDL and LDL environments, with more time spent at either endpoint than in the transition between them.

An alternative explanation of the bimodality could be that specific clathrate-like templates form, where the core could be more long-lived. Intermediate structures then appear in the interfacial region only as molecules attempt to attach to such templates before these collapse. We note that, although the life-time of individual H-bonds is of the order of a picosecond, there is currently no experimental information on the time-scale of *structural* transformations between HDL-like and LDL-like.

Assuming HDL to be chain-like[60], H-bond breaking and formation could simply correspond to exchanging which of the hydrogens that is H-bonded and which is free, which doesn't change the structure. Similarly, for a fully tetrahedrally H-bonded molecule in a tetrahedral environment, more than one bond must be simultaneously broken for the structure to break apart. There may thus exist dynamics on several time-scales in the liquid.

From a thermodynamic point of view, LDL is favored by the gain in energy from H-bond formation, but the high degree of local order that this enforces leads to loss of entropy. The HDL, on the other hand, is favored by entropy through more broken and distorted H-bonding. The loss in energy of directional H-bonds for HDL is compensated by packing the molecules closer, which gives more favorable van der Waals interaction[168]. However, maximal H-bond-order and structural disorder are mutually exclusive, which leads to a balance between the two and fluctuations in the liquid. Indeed, such a two-state picture, with a balance between competing thermodynamic contributions, has been shown to lead to a consistent description of liquid water across its phase diagram in terms of temperature and pressure[13,17-18,39-43].

Experimentally, HDL and LDL as pure liquids and the transition between them have been observed, but only in the ultraviscous regime[28]. The isothermal compressibility, $\kappa_T$, has been shown to exhibit a maximum at -44 °C (229 K) [8-9], which is what is expected in the one-phase region beyond a critical point[184]. However, the location of the proposed critical point lies inside what is called "No-man's land", which is the region bounded from above by the temperature ($T_H$) of homogeneous ice nucleation and from below by the temperature ($T_X$) of crystallization (see Fig. 8). In this "No-man's land" the liquid is extremely unstable and conversion to ice cannot be hindered. This makes locating a possible critical point in a liquid-liquid transition in this region extremely challenging. However, even the "virtual" existence of such a critical point would have a large influence also on water at conditions relevant for life. This would be a consequence of the two forms of liquid water, their coexistence and the mutual exclusivity of the thermodynamic contributions that favor each of them.

In Fig. 8 we illustrate in the form of a phase diagram in temperature and pressure the picture of liquid water that has emerged from recent and older data. At the bottom, at very low temperatures, we have the two forms, LDA and HDA, of amorphous ice that are known to interconvert in a first-order-like transition[26-27,185-186]. The coexistence line is assumed to continue into the "No-man's land" as indicated by the full black line in Fig. 8. This line is assumed to separate the pure HDL and LDL forms of the liquid. It should be noted that crystalline ice is far more stable at these temperatures and pressures than either of the liquid forms, so that, if one could still generate either liquid, it would quickly crystallize.

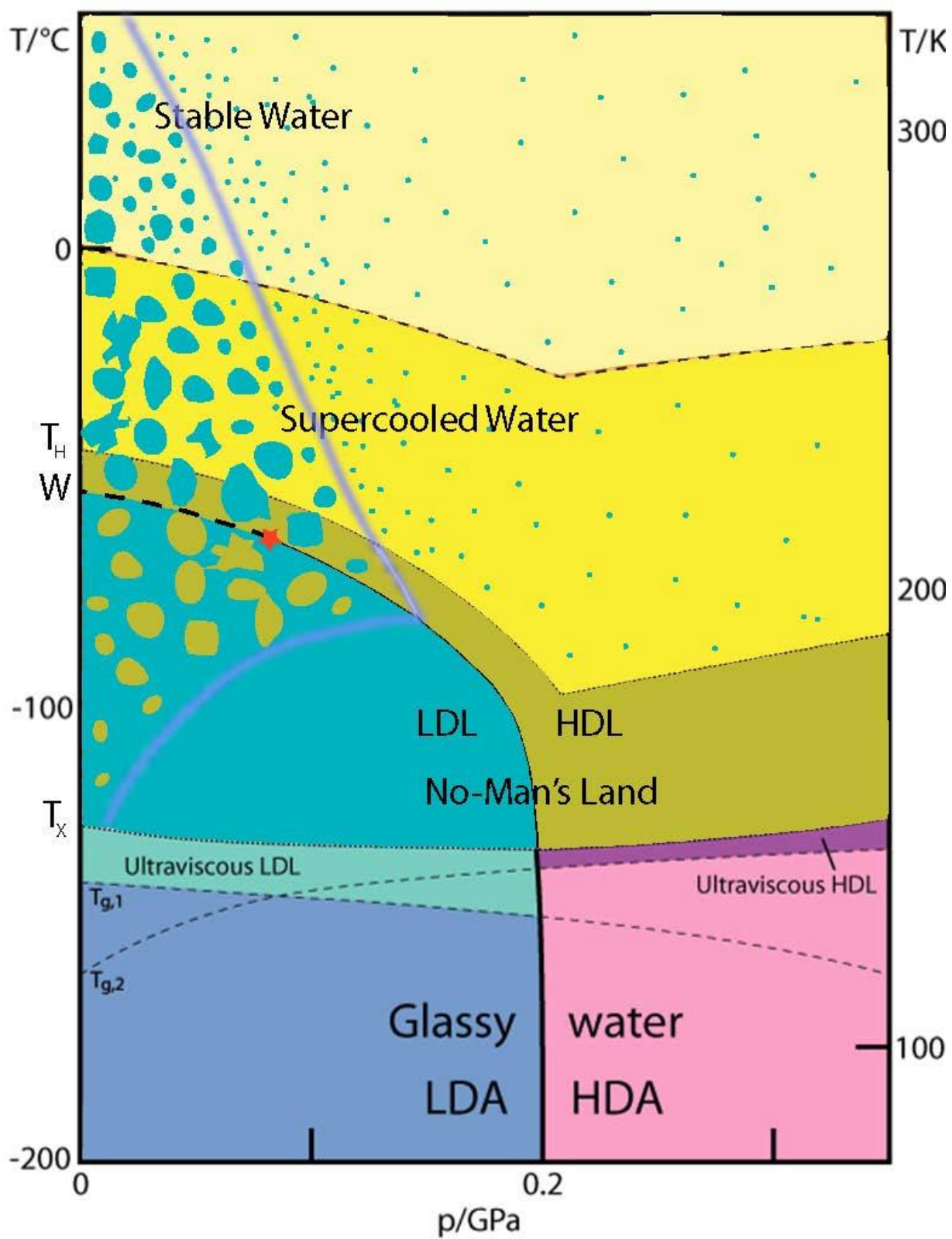

**Figure 8.** Illustrative phase diagram of liquid water in terms of pressure (x-axis) and temperature (y-axis). At the bottom there are the two forms of glassy water, LDA at low pressure and HDA at high pressure. $T_X$ indicates the crystallization limit as one heats the amorphous ices. The "No-man's land" region is subdivided into pure LDL

and pure HDL regions by the coexistence line (full black line) and terminated at the critical point (red star) beyond which the separation breaks down and large-scale fluctuations between the two forms appear. Below the Widom line (W, dashed), LDL (turquoise) dominates but with fluctuations into HDL regions (mustard), while above it, HDL dominates with fluctuations into LDL regions. The anomalous region, where correlated fluctuations occur and give deviations from simple liquid properties, lies in the funnel (hazy lines) extending from the region of the critical point. The farther away from the critical point one moves, the smaller the fluctuations become, as indicated by the size of the blobs. Outside the funnel, at higher temperatures, only local fluctuations occur in the otherwise HDL-dominated liquid (indicated by small blue dots on the yellow background). $T_H$ marks the temperature of homogeneous ice nucleation, below which the liquid will convert to ice, and 0 °C marks the equilibrium line between liquid and solid.

The red star in Fig. 8 indicates the proposed critical point terminating the coexistence line between HDL and LDL, while the dashed line, continuing the coexistence line beyond the critical point, indicates the Widom line (W). This is a line of maxima in correlation length (a measure of spatial extent of correlations), or isothermal compressibility ($\kappa_T$) or specific heat ($C_P$); there is thus more than one line of maxima beyond the critical point, but they all converge to the critical point[187].

The latter two properties, $\kappa_T$ and $C_P$, depend on fluctuations in, respectively, density and entropy, with the spatial extent of such fluctuations being greatest in the vicinity of the critical point (red star), as indicated by the size of the mustard and turquoise blobs in the illustration. It should be noted that, although this region beyond the critical point is a single-phase region where one can no longer speak of HDL and LDL, there will be fluctuations between the two structural species. The Widom line separates the region (turquoise) where LDL dominates, with HDL fluctuations, from that where HDL dominates, with LDL fluctuations (turquoise blobs). At the Widom line the ratio is 50:50[88], leading to maximum values of properties depending on fluctuations, such as $\kappa_T$ and $C_P$.

From the vicinity of the critical point there is a funnel-like region extending with structural fluctuations surviving also at temperatures and pressures far away from the critical point. This is indicated by the grey fuzzy lines extending as a funnel to lower and higher temperatures in the illustration and which delimit the region of correlated structural fluctuations between HDL and LDL. We define these lines as the onset of anomalous behavior of water properties, which thus occurs already around 50 °C for water at ambient pressure, as seen from the minimum of the isothermal compressibility (46 °C) and the associated structural changes evidenced by the temperature-dependent peaks in the O-O RDF[78,80]. This is the temperature and pressure region of life and we may thus be said to live within this funnel of water's structural fluctuations. This is a region where there is a fine balance between the strength of H-bonding being sufficient to survive thermal fluctuations in H-bonded proton-transport chains in enzymes, but not so strong as to inhibit flexibility in other aspects. The minimum in the heat capacity, $C_P$, is at 35 °C, which is close to the body temperature of large mammals like humans. Although the derivative of a function near its minimum is small, one may speculate that evolution has found an advantage in placing body temperature close to that at which it requires more energy to change it to higher or lower temperatures. This could be another aspect of "the funnel of life".

From the picture of ambient liquid water as being within "the funnel of life", we can now in very simple terms understand the anomalous properties of water. Since ambient liquid water is dominated by the HDL structure and ice is more like the LDL liquid, it is no wonder that the solid has less density and floats on the HDL-dominated liquid. We can view both HDL and LDL as normal liquids, which have their density increasing with decreasing temperature. At high temperatures water is dominated by HDL, but in the region of the funnel, water cannot decide which form is preferred. With more and more correlated LDL fluctuations appearing as water is cooled below 50 °C, it is clear that at some point the increasing appearance of lower-density structures should overcome the densification and a density maximum appear as the LDL fluctuations take over. For water this happens at 4 °C, which then is the temperature at the bottom of lakes and the oceans.

The high heat capacity of ambient liquid water can similarly be understood from the additional *structural* flexibility, added through the fluctuations into LDL-like structures in the HDL-dominated liquid. For normal, simple liquids, only random thermal fluctuations contribute, so the structural fluctuations in water add an extra dimension that increases the heat capacity and makes water a very efficient transporter of heat on our planet.

How does this picture relate to water as a solvent? The dipole moment of the water molecule makes water an excellent solvent for salts, but less so for non-polar solutes. The effects on the structure of water from dissolving salts (alkali fluorides, chlorides, bromides and iodides, *e.g.*, LiCl, NaCl, KCl, RbCl and CsCl) have been investigated using XAS[188]. The anions, except for the large iodide, were found to not significantly affect the water spectrum, while the pre-edge and main-edge in the spectra, associated with HDL-like structures, were enhanced by the cations. In conjunction with the loss of intensity at the post-edge (associated with LDL-like structures), this was interpreted as a change in the balance between HDL and LDL fluctuations, inhibiting the latter[188]. The effect on water structure from addition of salts was, furthermore, shown to be similar to heating water[23,37,188], which is known to break H-bonds, as well as to the effect of applying pressure[189], which would inhibit fluctuations into more voluminous LDL-like structures. The latter effect is seen directly from XRD on water at high pressures, which is found to reduce and eliminate the second shell in the O-O RDF that is associated with tetrahedrality[190].

Gases, such as $N_2$ and $O_2$, are small, non-polar molecules that do not interact strongly with the water molecules. We speculate that, in the choice between fitting into the dominating, compact HDL environment and a more spacious, open LDL-like environment, these molecules prefer the latter. Indeed, the solubility of these and similar gases increases strongly as water is cooled below 60 °C (333 K). This coincides with the appearance of correlated LDL-like fluctuations in the HDL-dominated liquid, and the solubility rises even more quickly as water is cooled further below ~20 °C (293 K). Furthermore, the solubility of, *e.g*., oxygen in seawater is lower than in fresh water, which might be connected to the effect of salts to reduce the occurrence of tetrahedral LDL-like fluctuations.

When considering how such a fluctuating picture of ambient water can affect efforts to purify water, the time-scale of structural fluctuations becomes essential. This is presently not known. If it is of the order of H-bond breaking and reformation, such fluctuations could be expected not to affect how water penetrates a separation membrane. If tetrahedral fluctuations are based on more long-lived structural templates, efficiency might be improved by working at higher temperatures, where the frequency of such fluctuations is reduced. This is highly speculative and would also depend on how water structure is affected by the presence of the membrane.

Figure 8 illustrates a picture of water, where two mutually exclusive contributions lead to two preferred arrangements in the liquid as close-packed HDL (favored by higher temperature and pressure) and the lower-density LDL. The proposed liquid-liquid critical point (red star in the figure) corresponds to the tipping point beyond which water cannot decide which form to prefer and fluctuations arise. These extend as a funnel-like region up to temperatures and pressures that are relevant to life and give rise to the numerous anomalous properties of the liquid that have caused scientists to refer to water as "the most anomalous liquid".

However, in the picture presented here, water is *not* a complicated liquid, but rather two simple liquids with a complicated relationship.

**ACKNOWLEDGEMENTS**

This chapter summarizes work from the literature and from experimental and modeling work from our group. Special thanks are due Anders Nilsson for many years of collaboration and discussions. Computational resources provided by the Swedish National Infrastructure for Computing (SNIC) at the HPC2N center are gratefully acknowledged.